\documentclass{article}
\usepackage{amssymb}
\usepackage{latexsym,amssymb}
\usepackage{graphics}
\usepackage{pstricks}
\usepackage{amsmath}
\usepackage{subfigure}
\usepackage{vmargin}
\usepackage{epsfig}

\begin{document}

\begin{titlepage}

\begin{flushright}
{CERN-PH-TH/2007-010}
\end{flushright}

\vskip 2cm

\begin{center}
\bf{Successful Yukawa structures in Warped Extra Dimensions}\\ 

\vskip 1.2cm

{J. I. Silva-Marcos \footnote{e-mail: {\tt juca@cftp.ist.utl.pt}} \\
{\it  CFTP, Departamento de F\'{\i}sica} \\
{\it  Instituto Superior T\'ecnico, Avenida Rovisco Pais, 1} \\
{\it 1049-001 Lisboa, Portugal}}
\end{center}

\vskip 1cm

\begin{abstract} 
For a RS model, with SM fields in the bulk and the Higgs boson on the TeV-brane, 
we suggest two specific structures for the Yukawa couplings, one based on a permutation symmetry 
and the other on the Universal Strength of Yukawa couplings hypothesis (USY). 
In USY, all Yukawa couplings have equal strength and the difference 
in the Yukawa structure lies in some complex phase. 
In both scenarios, all Yukawa couplings are of the same order of magnitude. 
Thus, the main features of the fermion hierarchies are explained through the RS geometrical mechanism, 
and not because some Yukawa coupling is extremely small.
We find that the RS model is particularly appropriate to incorporate the suggested Yukawa configurations. 
Indeed, the RS geometrical mechanism of fermion locations along the extra dimension, 
combined with the two Yukawa scenarios, 
reproduces all the present experimental data on fermion masses and mixing angles. 
It is quite remarkable that in the USY case, only two complex phases of definite value 
$\pm\frac{\pi}{2}$ are sufficient to generate the known neutrino mass differences, 
while at same time, permitting large leptonic mixing in agreement with experiment.
\end{abstract}

\vskip 1cm


\end{titlepage}

\section{Introduction}

\label{intro}

Lately, there has been great interest in extra dimension models. Extra
dimension models are inspired by string theory, which itself is based on the
existence of additional spatial dimensions \cite{Kaluza}. As known, string
theory \cite{String} is a main candidate for an all-including quantum theory
which allows for gravity, thus unifying all elementary particle interactions.

There exist several models of extra dimensions: universal extra dimension
models in which all Standard Model (SM) fields may propagate in extra
dimensions \cite{UED}, brane universe models in which the SM fields live in
our 3-dimensional spatial subspace \cite{Akama, Rubakov}, or intermediate
models in which only gauge bosons and the Higgs fields propagate in extra
dimensions while fermions are `stuck' at fixed points along these dimensions 
\cite{IntermA}. Amongst the brane universe models two different scenarios
have attracted much attention: one suggested by Arkani-Hamed, Dimopoulos and
Dvali (ADD) with large flat extra dimensions \cite{ADD}, and the other by
Randall and Sundrum (RS) with a single small warped extra dimension \cite%
{Gog,RS}\footnote{%
See also \cite{+-+A} for extensions of the RS model.}.

Extra dimensional models have some advantages over supersymmetric theories 
\cite{SUSY}. Besides the fact that they lead to the unification of \ the
gauge couplings, either at high $10^{16}$ GeV scales for small warped extra
dimension models \cite{UNI-RS}, or at the lower TeV scales for large flat
extra dimension models \cite{UNI}, they (the ADD and RS brane models) also
address the long standing puzzle of the gauge hierarchy problem, i.e. the
huge discrepancy between the gravitational scale and the electroweak scale.
Furthermore, there is a viable Kaluza-Klein WIMP candidate for the dark
matter of the universe \cite{KKrelic}. In addition, extra dimensional models
explain the large mass hierarchy of the different types and generations of
the SM fermions through a geometrical mechanism. In both, the ADD models 
\cite{AS}-\cite{Klap} and in the RS models \cite{RSloc,RSmany}, the SM
fermions have different localizations along extra dimension(s) which depend
on the type and the flavour of the fermions. This mechanism does not rely on
the presence of any new symmetry in the short-distance theory, as in the
case of the conventional Froggatt-Nielsen mechanism, which introduces a
`flavor symmetry' \cite{FN}.

One finds in the literature many higher-dimensional models which have been
proposed for generating the SM fermion mass hierarchy \cite{others1}. E.g.
several frameworks explain the lightness of neutrinos relatively to other SM
fermions within the ADD \cite{VolMesPow1}, \ the RS \cite%
{GNeubertA,GNeubertB} or RS extensions \cite{mypaper}.

We concentrate on a RS model, which in contrast with the ADD, does not need
of any new energy scale in the fundamental theory. In this RS model, the SM
fermion mass hierarchy arises mainly from the dependence on the location,
within the warped geometry, of each (type and family) fermion. However, to
obtain the correct quark and lepton (masses and) mixing, one must have
appropriate Yukawa coupling structures, because, even if the Yukawa
couplings do not induce the main features of the fermion mass hierarchy
(coming thus from the geometrical mechanism), the structure of the Yukawa
texture is still of crucial significance, especially for the mixing. Within
the context of the different SM fermion locations along a warped extra
dimension, diverse cases have been studied \cite{Eigencite1}, e.g. models
with Majorana neutrinos \cite{HSHHLL}, or models with Dirac neutrinos where
the charged lepton Yukawa couplings are assumed to be diagonal \cite{HSpre},
or models with small Yukawa fluctuations \cite{Eigen, Eigen1}. In all these
cases, however, the Yukawa couplings did not come from any principle or
symmetry, but were chosen in such a way as to merely justify the data.

In this paper, we consider Dirac fermions and give two specific structures
for the fermion Yukawa coupling texture, which seem to be particularly
appropriate for the RS model, and which account for the experimental data on
masses and mixings of all fermions, leptons and quarks. We focus\ on a
permutation symmetry structure and on the Universal Strength for Yukawa
couplings (USY) \cite{USY}. In USY, all Yukawa couplings have equal strength
and the difference in the Yukawa structure lies in some complex phase. In
both scenarios, all Yukawa couplings are of the same order of magnitude. The
main features of the fermion mass hierarchy comes thus from the geometrical
mechanism, but the correct mixings and masses are obtained from Yukawa
textures which have a definite structure (USY) or a (permutation) symmetry .

In Section \ref{frame}, we describe the RS model and the geometrical
mechanism, which allows for different localizations along the warped extra
dimension, together with the resulting fermion mass matrices in 4D. In
Section \ref{const}, we make a short review of the experimental constraints
on the RS scenario parameters. In Section \ref{real}, we concentrate on the
structure of the Yukawa couplings and their importance for the mixing. Two
appropriate structures for the Yukawa couplings in the RS scenario are
introduced: a permutation symmetry and USY. Then, in Section \ref{pred},
predictions are given for fermion masses and mixing angles based on the two
proposed texture scenarios. Finally, we conclude in Section \ref{conclu}.

\section{Theoretical framework}

\label{frame}

\subsection{The RS geometrical scenario}

\label{RSgeom}

The RS scenario consists of a 5-dimensional theory, where the extra spatial
dimension, parameterized by $y$, is compactified on a $S^{1}/\mathbb{Z}_{2}$
orbifold of radius $R$, such that $-\pi R\leq y\leq \pi R$. Gravity
propagates in the bulk and the extra spatial dimension is bordered by two
3-branes with tensions $\Lambda _{y=0}$ and $\Lambda _{y=\pi R}$ (vacuum
energy densities) tuned in such way that, 
\begin{equation}
\Lambda _{(y=0)}=-\Lambda _{(y=\pi R_{c})}=-\Lambda /k=24kM_{5}^{3},
\label{RStensions}
\end{equation}%
where $\Lambda $ is the bulk cosmological constant, $M_{5}$ the fundamental
5-dimensional gravity scale and $k$ a characteristic energy scale (typically
of the order $M_{Pl}$,). One finds a solution to the 5-dimensional
Einstein's equations which respects 4-dimensional Poincar\'{e} invariance.
This solution corresponds to a zero mode of the graviton localized on the
positive tension brane (the 3-brane at $y=0$) and the non-factorizable
metric, 
\begin{equation}
ds^{2}=e^{-2\sigma (y)}\eta _{\mu \nu }dx^{\mu }dx^{\nu }+dy^{2},\ %
\mbox{with}\ \sigma (y)=k|y|,  \label{RSmetric}
\end{equation}%
where $x^{\mu }$ denotes the coordinates of the familiar 4 dimensions and $%
\eta _{\mu \nu }=diag(-1,1,1,1)$ the flat metric. The bulk geometry
associated to the metric in Eq. (\ref{RSmetric}) is a slice of
Anti-de-Sitter ($AdS_{5}$) space with curvature radius $1/k$.

From the fluctuations on the metric of Eq. (\ref{RSmetric}), one derives
(integrating over $y$) an expression for the effective 4-dimensional gravity
scale as a function of the three fundamental RS parameters $k$, $R$ and $%
M_{5}$: 
\begin{equation}
M_{Pl}^{2}=\frac{M_{5}^{3}}{k}(1-e^{-2\pi kR}).  \label{RSkrelat}
\end{equation}

On the Planck-brane (the 3-brane at $y=0$), the gravity scale is equal to
the (reduced) Planck mass: $M_{Pl}=1/\sqrt{8\pi G_{N}}=2.44\ 10^{18}%
\mbox{GeV}$ ($G_{N}\equiv $ Newton's constant), while on the TeV-brane (the
3-brane at $y=\pi R$), the gravity scale is affected by the exponential
\textquotedblleft warp\textquotedblright\ factor $w=e^{-\pi kR_{c}}$and
becomes much smaller: 
\begin{equation}
M_{\star }=w\ M_{Pl},  \label{RSratio}
\end{equation}%
Clearly, for a small extra dimension with radius $R\simeq 11/k$, with $k$ of
order $M_{Pl}$, one has $w\sim 10^{-15}$ and thus $M_{\star }=\mathcal{O}%
(1)\ \mbox{TeV}$. Hence, the gravity scale $M_{\star }$ on the TeV-brane can
be of the same order of magnitude as the electroweak symmetry breaking scale.

In addition, on finds a solution for the gauge hierarchy problem: if the SM
Higgs boson is confined on the TeV-brane, it feels a cut-off at $M_{\star }=%
\mathcal{O}(1)\ \mbox{TeV}$ which guarantees the stability of Higgs mass
with respect to divergent quantum corrections.

\subsection{The fermion mass matrices in 4D}

\label{effectiveM}

In this RS scenario, the SM fermion mass hierarchy is generated through a
geometrical mechanism: the fermions reside in the bulk\footnote{%
Gauge coupling unification and 5-dimensional gauge invariance require the SM
gauge bosons to live in the bulk \cite{HRizzo, hep-ph/9912498}. A
Kaluza-Klein WIMP candidate (within the RS model) also requires bulk
fermions. See \cite{GNeubertA}, for the behavior of fermions in the bulk.}
and the SM (zero mode) fermions are given different localizations along the
warped extra dimension. Each type of SM fermion field $\Psi _{i}$ ($i$ is
the flavor index) is coupled to a mass $m_{i}$ in the 5-dimensional
fundamental theory:

\begin{equation}
\ m_{i}\bar{\Psi}_{i}\Psi _{i},  \label{mass}
\end{equation}%
To have different localizations for the zero mode fermions, one must have a
non-trivial dependence of the $m_{i}$ on the fifth dimension, i.e. the $%
m_{i} $, which could be the vacuum expectation values of some scalar fields,
must have a `(multi-)kink' profile \cite{Rubakov,Rebbi}. An attractive
possibility is a parameterization of the masses as \cite{Tamvakis}, 
\begin{equation}
m_{i}=c_{i}\ \frac{d\sigma (y)}{dy}=\pm \ c_{i}\ k,  \label{VEV}
\end{equation}%
where $\sigma (y)$ is defined in Eq.(\ref{RSmetric}) and the $c_{i}$ are
dimensionless parameters. This parameterization is compatible with the $%
\mathbb{Z}_{2}$ symmetry ($y\rightarrow -y$) of the $S^{1}/\mathbb{Z}_{2}$
orbifold: the masses are odd under the $\mathbb{Z}_{2}$ transformation.
Defining the fermion parity with $\Psi _{\pm }(-y)=\pm \gamma _{5}\Psi _{\pm
}(y)$, one finds that the product $\bar{\Psi}_{\pm }^{i}\Psi _{\pm }^{i}$ is
also odd under the $\mathbb{Z}_{2}$ transformations, and thus, the whole
mass term in Eq. (\ref{mass}) is even.

The equation of motion for the 5-dimensional fermion fields $\Psi _{i}$
includes the mass term in Eq. (\ref{mass}) and the $\Psi _{i}$ decompose as, 
\begin{equation}
\Psi _{i}(x^{\mu },y)=\frac{1}{\sqrt{2\pi R_{c}}}\sum_{n=0}^{\infty }\psi
_{i}^{(n)}(x^{\mu })\ f_{n}^{i}(y),  \label{dec}
\end{equation}%
where $n$ labels the tower of Kaluza-Klein excitations. The zero mode wave
function admits the following solution along the extra dimension \cite%
{RSloc,GNeubertA}, with normalization factor $N_{0}^{i}$: 
\begin{equation}
\begin{array}{ccc}
f_{0}^{i}(y)=\frac{e^{(2-c_{i})\sigma (y)}}{N_{0}^{i}} & \quad ;\quad & 
N_{0}^{i}=\sqrt{\frac{e^{\pi kR(1-2c_{i})}-1}{2\pi kR(1-2c_{i})}}.%
\end{array}
\label{0mode}
\end{equation}%
From Eq.(\ref{0mode}) one finds that the zero mode of a fermion is more
localized towards the Planck-brane if $c_{i}$ increases. Subsequently, the
zero mode of the fermion is more localized towards the TeV brane if $c_{i}$
decreases. We shall see that localization differences are important to
generate the fermion mass hierarchies.

The SM fermions acquire a Dirac mass through the coupling to a Higgs field.
After spontaneous symmetry breaking, and integrating out the extra dimension
one obtains: 
\begin{equation}
\int d^{4}x\int dy\ \sqrt{G}\ \left( \lambda _{ij}^{(5)}\ H\ \bar{\Psi}%
_{+i}\Psi _{-j}+h.c.\right) =\int d^{4}x\ M_{ij}\ \bar{\psi}_{Li}^{(0)}\psi
_{Rj}^{(0)}+h.c.+\dots ,  \label{Yuk}
\end{equation}%
where $G=e^{-8\sigma (y)}$ is the determinant of the RS metric. The $\lambda
_{ij}^{(5)}$ are the 5-dimensional Yukawa coupling constants and the dots
stand for the mass terms of the Kaluza-Klein (KK) excited modes. The
effective 4-dimensional mass matrix is given by the integral:%
\begin{equation}
M_{ij}=\int dy\ \sqrt{G}\ \frac{\lambda _{ij}^{(5)}}{2\pi R}\ H(y)\
f_{0}^{i}(y)\ f_{0}^{j}(y).  \label{MassMatrix}
\end{equation}%
Following the motivation from the equation of motion for a bulk scalar field 
\cite{GWise}, we assume an exponential form for the Higgs field: 
\begin{equation}
H(y)=H_{0}\ e^{4k(|y|-\pi R)},  \label{Hprofile}
\end{equation}%
which is shape-peaked at the TeV-brane. As mentioned in Section \ref{RSgeom}%
, this is a crucial ingredient in obtaining a solution for the gauge
hierarchy problem. The amplitude $H_{0}$ can be expressed in terms of the $%
W^{\pm }$ boson mass, $kR$ and the 5-dimensional weak gauge coupling
constant $g^{(5)}$.

After integration of Eq. (\ref{MassMatrix}), one obtains an analytical
expression for the fermion mass matrix, 
\begin{equation}
M_{ij}=\ \lambda _{ij}\ M_{kR}\ \left( \frac{e^{\pi kR(4-c_{i}-c_{j})}-1}{%
4-c_{i}-c_{j}}\right) \left( \frac{(1-2c_{i})(1-2c_{j})}{(e^{\pi
kR(1-2c_{i})}-1)\ (e^{\pi kR(1-2c_{j})}-1)}\right) ^{\frac{1}{2}}\ .
\label{analytica}
\end{equation}%
where%
\begin{equation*}
M_{kR}=\frac{1}{2}M_{W}\sqrt{\frac{6\pi kR}{e^{6\pi kR}-1}}
\end{equation*}%
It is assumed that $\lambda _{ij}^{(5)}=\lambda _{ij}\ g^{(5)}$.
Consequently, there is no $g^{(5)}$ dependence in the expression of Eq. (\ref%
{analytica}), which is exactly compensated by the $g^{(5)}$ dependence in
the amplitude $H_{0}$. For each type of fermion, one obtains a different
(Dirac) mass matrix , 
\begin{equation}
M_{ij}^{a}=M_{ij}^{a}(kR,\lambda _{ij}^{a},c_{i}^{A},c_{j}^{a})\ \ 
\label{depend}
\end{equation}%
In this notation, $a=l,\nu $ together with $A=L$ for the leptons, while for
the quarks, $a=u,d$ with $A=Q$. The $c_{i}^{L}$ parameterize the
5-dimensional masses of the $SU(2)_{L}$ doublets of the leptons while the $%
c_{j}^{l}$, $c_{j}^{\nu }$ parameterize the 5-dimensional masses of the
right handed charged leptons or neutrinos. The $c_{i}^{Q}$ and $c_{j}^{u}$, $%
c_{j}^{d}$ apply to the quarks. For each type of fermion we have different
Yukawa couplings: $\lambda _{ij}^{l}$, $\lambda _{ij}^{\nu }$ \ for the
leptons and $\lambda _{ij}^{u}$, $\lambda _{ij}^{d}$ for\ the quarks.

From the exponential character of the functions in the expression of $M_{ij}$
in Eq.(\ref{analytica}), it is clear that the fermion masses can differ
greatly (spanning several orders of magnitude) for each flavor $i,j$,
depending on the values of $c_{i}$, $c_{j}$. This dependence resulted from
the overlap between Higgs profile $H(y)$ and zero mode fermion wave function 
$f_{0}^{i,j}(y)$ which also varies with flavor through the $c_{i}$, $c_{j}$
parameters (see Eq.(\ref{0mode}).

\section{Experimental constraints and Parameter space}

\label{const}

\label{large}

Next, we describe several constraints on the parameters of the model. The
parameter space includes the scales $k$, $M_{5}$ and the radius of the extra
dimension $R$, together with the $c_{i}^{L}$, $c_{j}^{l}$, $c_{j}^{\nu }$
for the leptons and $c_{i}^{Q}$, $c_{j}^{u}$, $c_{j}^{d}$ for the quarks,
which parametrize the 5-dimensional masses.

The bulk curvature must be small compared to the higher-dimensional gravity
scale, i.e. $k\leq M_{5}$ if one is to trust the RS solution for the metric (%
\textit{c.f.} Eq.(\ref{RSmetric})) \cite{RS}. For the limiting situation (as
in \cite{HSpre,HSquark,HSfv}) where $k=M_{5}$, there is just one energy
scale and $M_{5}=M_{Pl}$. This results from Eq.(\ref{RSkrelat}) and the fact
that one must have $kR\simeq 11$, to have a sufficiently small wrap factor
for gravity to be of the order of the TeV scale on the TeV brane. We shall
keep $k=\mathcal{O}(M_{5})$. Indeed, for $kR=10.83$, consistent with a
5-dimensional gravity scale (on the TeV-brane) of $M_{\star }=4\mbox{TeV}$ (%
\textit{c.f.} Eq.(\ref{RSratio})) and in agreement with the solution
proposed for the gauge hierarchy problem, a typical mass value for $m_{KK}$
is obtained\footnote{%
The mass of first gauge boson KK excitation is given by $m_{KK}=2.45\ k\
e^{-\pi kR_{c}}$ in the RS model \cite{EWboundD}.\label{mKK}} of $m_{KK}=1%
\mbox{TeV}$ if one chooses $k=0.1M_{Pl}$. Once the $k$ and $R$ parameter
values are known, the $M_{5}$ value is fixed by Eq.(\ref{RSkrelat}). For
these $k$ and $R$ values, $M_{5}=1.13\ 10^{18}\mbox{GeV}$. Thus, the two
values of the fundamental energy scales $k$ and $M_{5}$ in the RS model are
quite close. However from Eq.(\ref{RSkrelat}), it is clear that any choice
of $k$ such that $0.1M_{Pl}<k<10M_{Pl}$ results in fundamental energy scales 
$k$ and $M_{5}$ which are close. This enables us to choose other $kR$ values
(keeping $kR\simeq 11$). Here, we do not focus on the precise value of $kR$.
Indeed, slightly different $c_{i}$ parameters can be found, in agreement
with the permutation and USY setup of this paper, if one cares to choose
other $kR$ or $m_{KK}$ values.

Precision electroweak (EW) data constrain the RS model \cite{Eigen1} \cite%
{EWboundA}-\cite{EWboundE}. All fields have KK excited states and these lead
to new contributions to physical quantities \cite{Eigencite2}. Here, we name
a few. E.g. the mixing between the top quark and its KK excited states
contributes to the $\rho $ parameter, which might exceed the bound set by
precision EW measurements \cite{EWboundA,hep-ph/0204002}. However, if one
chooses certain localization configurations for quark fields (i.e. certain
values for $c_{i}$ quark parameters), some of these problems may be
circumvented \cite{Eigen1}.

The mixing between the EW gauge bosons and their KK modes also induce
deviations for some precision EW observables. These go typically as $%
(m_{W}/m_{KK})^{2}$, where $m_{KK}=m_{KK}^{(1)}(W^{\pm })$ is the mass of
first KK excitation of the $W$ gauge boson \footnote{%
The difference between $m_{KK}^{(1)}(W^{\pm })$ and $m_{KK}^{(1)}(Z^{0})$ is
insignificant in the RS model.}. Deviations to the weak gauge boson masses
and the W boson coupling to fermions on the Planck-brane (TeV-brane) lead to
experimental bounds on $m_{KK}$. These depend on the localization of the SM
fermion fields in the bulk (i.e. the values of mass parameters $c_{i}$ for
SM fermions). Typically, one finds $m_{KK}\gtrsim 4\mbox{TeV}$ \cite%
{EWboundB}, but lower values, down to $m_{KK}=1\mbox{TeV}$, are allowed for
certain $c_{i}$. A global analysis of these constraints can be found in \cite%
{Eigen, Eigen1}.

Experimental bounds on Flavor Changing (FC) processes constrain the RS
model, since significant FC effects can be generated with bulk SM fields 
\cite{HSpre,HSfv,FCbound}. The exchange of heavy lepton KK excitations leads
to deviations from unitarity for the leptonic mixing matrix, which
contributes to FC processes like lepton decays: $\mu \rightarrow e\gamma $, $%
\tau \rightarrow \mu \gamma $ and $\tau \rightarrow e\gamma $. In \cite%
{Eigen1} we showed that, even for a small value such as $m_{KK}=1\mbox{TeV}$%
, there exist suitable values for the mass parameters $c_{i}$ for which the
branching ratios for these three rare decays are below their experimental
upper limit.

With regard to the values of the $c_{i}$ parameters, we assume $%
|c_{i}|\approx \mathcal{O}(1)$. Thus, the natural values of 5-dimensional
masses $m_{i}$ (\textit{c.f.} Eq.(\ref{VEV})) appearing in the original
action are of the same order of magnitude as the fundamental scale of the RS
model, namely the bulk gravity scale $M_{5}$, and we avoid the introduction
of new energy scales in the theory. For $k=\mathcal{O}(M_{5})$, the absolute
values of the $c_{i}$ parameters should be of the order of unity. Typically,
here we take,%
\begin{equation}
|c_{i}|\lessapprox 5.  \label{cB}
\end{equation}

A precise analysis on the possible values and restrictions on $m_{KK}$ and
the allowed values for the $c_{i}$ is outside the scope of this paper. The $%
c_{i}$ values (or the exact $kR$ value) that we shall use in this paper are
not screened for their ElectroWeak precision compatibility, as we merely
wish to illustrate that it is possible, in a first anylisis, to incorporate
the data on masses and mixings. We refer to \cite{Eigen, Eigen1}, for more
details on these, as well as many other effects caused by the diverse KK
states.

\section{Yukawa couplings and mass matrices}

\label{real}

\subsection{The structure of the mass matrices}

Next, we give an analysis of the fermion mass matrices $M_{ij}^{a}$ which
result from our RS scenario. From the expression for the $M_{ij}$ given in
Eq.(\ref{depend}), we find that, to a good approximation, the $M_{ij}$ can
be written as 
\begin{equation}
M_{ij}^{a}=g_{_{i}}(c_{i}^{A})\cdot \kappa _{ij}^{a}\ \ \cdot \widehat{g}%
_{_{j}}(c_{j}^{a})  \label{sep}
\end{equation}%
This splitting of $M_{ij}$ is valid in large regions of the parameter space
spanned by $c_{i}^{A}$, $c_{j}^{a}$. The $g_{_{i}},\ \widehat{g}_{_{j}}$ are
suitable functions for certain regions. E.g. for the region $1/2<c_{i}^{A}$, 
$c_{i}^{a}<3/4$, the functions are equal, $g_{_{i}}=\ \widehat{g}_{_{j}}$ $%
=g $, with 
\begin{equation}
g(x)=\lambda \sqrt{\frac{(2x-1)}{2-x}}\ e^{\pi kR_{c}(2-x)}.  \label{gfunc}
\end{equation}%
where $\lambda $ is some parameter dependent on $kR$ and $M_{W}$ (but
irrelevant for the following).

The splitting-structure of $M_{ij}$, however, has important consequences. To
see this, take as an example all $\kappa _{ij}=1$. Then, from the structure
in Eq. (\ref{sep}), we obtain, e.g. for the neutrinos and the charged
leptons, the following expressions for the mass matrices: 
\begin{equation}
\begin{array}{lll}
\begin{array}{c}
M_{\nu }=D_{L}\cdot \Delta \cdot D_{\nu } \\ 
\\ 
M_{l}=D_{L}\cdot \Delta \cdot D_{l}%
\end{array}
& \ ;\ \ \ \  & \Delta =\left[ 
\begin{array}{ccc}
1 & 1 & 1 \\ 
1 & 1 & 1 \\ 
1 & 1 & 1%
\end{array}%
\right]%
\end{array}
\label{mnch}
\end{equation}%
where $D_{L,l,\nu }=\mathrm{diag}(a_{1,}a_{2},a_{3})_{L,l,\nu }$, and the $a$%
's are obtained from the $g_{_{i}}$ and $\widehat{g}_{_{j}}$ functions in
Eq. (\ref{sep}). In this approximation, only the tau and one neutrino
eigenstate have mass. However, the point is, that the resulting squared
matrices $H_{\nu }=$ $M_{\nu }M_{\nu }^{\dagger }$ and $H_{l}=$ $%
M_{l}M_{l}^{\dagger }$ are of the same form 
\begin{equation}
\begin{array}{lll}
\begin{array}{c}
H_{\nu }=\rho _{\nu }\ D_{L}\cdot \Delta \cdot D_{L} \\ 
\\ 
H_{l}=\rho _{l}\ D_{L}\cdot \Delta \cdot D_{L}%
\end{array}
& \ ;\ \ \ \  & 
\begin{array}{c}
\rho _{\nu }=\sqrt{a_{\nu _{1}}^{2}+a_{\nu _{2}}^{2}+a_{\nu _{3}}^{2}} \\ 
\\ 
\rho _{l}=\sqrt{a_{l_{1}}^{2}+a_{l_{2}}^{2}+a_{l_{3}}^{2}}%
\end{array}%
\end{array}
\label{hnch}
\end{equation}%
Thus, the matrices $V_{\nu }$ and $V_{l}$, which diagonalize respectively $%
H_{\nu }$ and $H_{l}$, are equal; there is no mixing: $V_{MNS}=V_{l}^{%
\dagger }V_{\nu }={1\>\!\!\!\mathrm{I}}$, and although small deviations from 
$\kappa _{ij}^{l,\nu }=1$ may be sufficient to generate the masses of all
other charged leptons and neutrinos, this scenario only leads to small
deviations from $V_{MNS}={1\>\!\!\!\mathrm{I}}$, for the mixing.

It is clear, in RS scenarios, even if the Yukawa couplings do not induce the
main features of the fermion mass hierarchy (coming thus from the
geometrical mechanism described), the structure of the Yukawa couplings is
of crucial significance for the mixing, and at least for the leptons, some
of the $\kappa _{ij}^{l,\nu }$ must be very different from$\ $each other.

\subsection{Successful scenarios for the Yukawa structure}

In this subsection, we give two possible structures for the Yukawa
couplings, which are compatible with our RS scenario. We focus\ a
permutation symmetry structure and the Universal Strength for Yukawa
couplings (USY) texture \cite{USY}. In both scenarios, all Yukawa couplings
are of the same order of magnitude. Thus, the main features of the fermion
hierarchies are explained through the RS geometrical mechanism, and not
because some Yukawa coupling is extremely small. Explicitly, we choose $%
1/2<|k_{ij}|<2$.

(A) \textbf{Permutations}

Imposing the following permutation structure on the fermion fields:%
\begin{equation}
\begin{array}{c}
\Psi _{1}^{A}\longrightarrow \Psi _{2}^{A}\ ;\ \ \ \ \Psi
_{2}^{A}\longrightarrow \Psi _{3}^{A}\ ;\ \ \ \ \Psi _{3}^{A}\longrightarrow
\Psi _{1}^{A} \\ 
\\ 
\Psi _{1}^{a}\longrightarrow \Psi _{2}^{a}\ ;\ \ \ \ \Psi
_{2}^{a}\longrightarrow \Psi _{3}^{a}\ ;\ \ \ \ \Psi _{3}^{a}\longrightarrow
\Psi _{1}^{a}%
\end{array}
\label{permutation}
\end{equation}%
(remember the notation: $a=l,\nu $ when $A=L$ for the leptons, while for the
quarks, $a=u,d$ with $A=Q$). Then, all Yukawa couplings have the following
structure:%
\begin{equation}
k=\left[ 
\begin{array}{ccc}
1 & A & B \\ 
B & 1 & A \\ 
A & B & 1%
\end{array}%
\right]  \label{yuks3}
\end{equation}%
In addition to this, the $c_{i}$'s and $m_{i}$ 's in Eq. (\ref{VEV}) of each
type of field would be equal. However, at this point, we break the symmetry
by some mechanism, e.g. by having different vacuum expectation values for
the scalar fields which give rise to the $m_{i}$. Thus, we take $%
c_{i}^{A}\neq c_{j}^{A}$ and $c_{i}^{a}\neq c_{j}^{a}$ for $i\neq j$. It is
known that in the SM, Yukawa couplings of the type in Eq. (\ref{yuks3}) lead
to masses and mixings which are non realistic.

\bigskip

(B) \textbf{USY}

Another possible structure for the Yukawa couplings is the so called
Universal Strength for Yukawa couplings. In USY, all Yukawa couplings have
equal strength, $|k_{ij}|=1$. The difference in the Yukawa structure lies in
some complex phase:%
\begin{equation}
k_{ij}=\left[ e^{\theta _{ij}}\right]  \label{usy}
\end{equation}%
We shall see that this structure is particularly useful for the neutrinos,
as a minimum of two (indeed, curious) complex phases is already sufficient
to generate the known neutrino mass differences.

\section{Results on fermion masses and mixing}

\label{pred}

\label{fitdata}

\subsubsection{The experimental data}

Next, we present the data on fermion masses and mixings. In principle, we
consider running masses at the cutoff energy scale of the effective
4-dimensional theory, i.e. the TeV range. This scale is also of the order of
the electroweak symmetry breaking scale. The predictions for the fermion
masses are fitted with the experimental mass values taken at the pole \cite%
{PDG}. We allow for an uncertainty of $5\%$ in order to take into account
the effect of the renormalization group from the pole mass scale up to the
TeV cutoff scale. This is only a few percent \cite{hep-ph/9912265}. In
accordance with this uncertainty, we do not determine experimental values
with high accuracy. E.g. we take the experimental data on neutrino masses
and leptonic mixing angles at the $4\sigma $ level \cite{Valle}.

With respect to the neutrinos, a general three-flavor fit to the current
world's global neutrino data sample has been performed in \cite{Valle},
which includes the results from solar, atmospheric, reactor (KamLAND and
CHOOZ) and accelerator (K2K) experiments. The values for oscillation
parameters obtained in this analysis at the $4\sigma $ level are contained
in the intervals: 
\begin{equation*}
6.8\leq \Delta m_{21}^{2}\leq 9.3\ \ \ [10^{-5}\mbox{eV}^{2}],
\end{equation*}%
\begin{equation}
1.1\leq \Delta m_{31}^{2}\leq 3.7\ \ \ [10^{-3}\mbox{eV}^{2}],
\label{4SigmaDataA}
\end{equation}%
where $\Delta m_{21}^{2}\equiv m_{\nu _{2}}^{2}-m_{\nu _{1}}^{2}$ and $%
\Delta m_{31}^{2}\equiv m_{\nu _{3}}^{2}-m_{\nu _{1}}^{2}$ are the
differences of squared neutrino mass eigenvalues, and, 
\begin{equation*}
0.21\leq \sin ^{2}\theta _{12}\leq 0.41,
\end{equation*}%
\begin{equation*}
0.30\leq \sin ^{2}\theta _{23}\leq 0.72,
\end{equation*}%
\begin{equation}
\sin ^{2}\theta _{13}\leq 0.073,  \label{4SigmaDataB}
\end{equation}%
where $\theta _{12}$, $\theta _{23}$ and $\theta _{13}$ are the three mixing
angles from the standard parameterization of the leptonic mixing matrix. The
data on tritium beta decay \cite{Farzan} provided by the Mainz and Troitsk 
\cite{Troitsk} experiments give some experimental limits on the effective
neutrino mass at $95\%\ C.L.$, 
\begin{equation*}
m_{\beta }\leq 2.2\ \mbox{eV}\ \ \ \mbox{[Mainz]},
\end{equation*}%
\begin{equation}
m_{\beta }\leq 2.5\ \mbox{eV}\ \ \ \mbox{[Troitsk]},  \label{mbetaLIM}
\end{equation}%
with $m_{\beta }$ defined by, $m_{\beta
}^{2}=\sum_{i=1}^{3}|V_{ei}|^{2}m_{\nu _{i}}^{2}$, where $V_{ei}$ is
leptonic mixing matrix..

At $M_{Z}$, the renormalized charged lepton masses are \cite{Fus} 
\begin{equation}
\begin{array}{l}
m_{e}=0.48684727\pm 1.4\ 10^{-7}\ \mbox{MeV} \\ 
m_{\mu }=102.75138\pm 3.3\ 10^{-4}\ \mbox{MeV} \\ 
m_{\tau }=1.74669_{-0.00027}^{+0.00030}\ \mbox{GeV}%
\end{array}
\label{dataL}
\end{equation}%
The quark masses and CKM matrix parameters at $M_{Z}$ are give by \cite{Fus}%
, 
\begin{equation}
\begin{array}{ll}
m_{d}=4.69_{-0.66}^{+0.60}\ \mbox{MeV}\ ; & m_{u}=2.33_{-0.45}^{+0.42}\ %
\mbox{MeV} \\ 
m_{s}=93.4_{-13.0}^{+11.8}\ \mbox{MeV}\ ; & m_{c}=677_{-61}^{+56}\ \mbox{MeV}
\\ 
m_{b}=3.00\pm 0.11\ \mbox{GeV}\ ; & m_{t}=181\pm 13\ \mbox{GeV}%
\end{array}
\label{dataQ}
\end{equation}%
and by 
\begin{equation}
\begin{array}{rcl}
|V_{us}| & = & 0.2205\pm 0.0018 \\ 
|V_{cb}| & = & 0.0373\pm 0.0018 \\ 
|V_{ub}/V_{cb}| & = & 0.08\pm 0.02.%
\end{array}
\label{dataV}
\end{equation}

\subsection{Results for two Yukawa scenarios}

\label{finalparam}

Next, we give 4 examples of the parameter values for the $c_{i}$'s and
Yukawa couplings for the two scenarios (permutations and USY), which
together with the RS mechanism described, reproduce the known experimental
data on fermion masses and mixings.

(A) \textbf{Permutations}

Taking the following values for the $c_{i}^{L}$, $c_{j}^{l}$, $c_{j}^{\nu }$
and the permutation matrix parameters $A^{l,\upsilon },B^{l,\upsilon }$ in
Eq. (\ref{yuks3}) 
\begin{equation}
\begin{array}{lll}
c_{1}^{L}=0.2 & c_{1}^{l}=0.862 & c_{1}^{\upsilon }=1.485 \\ 
c_{2}^{L}=0.2 & c_{2}^{l}=0.698 & c_{2}^{\upsilon }=1.725 \\ 
c_{3}^{L}=0.2 & c_{3}^{l}=0.647 & c_{3}^{\upsilon }=1.585%
\end{array}%
\ ;\ \ \ 
\begin{array}{cc}
A^{l}=1.5 & A^{\upsilon }=-0.6 \\ 
B^{l}=1.5 & B^{\upsilon }=1.7%
\end{array}
\label{lep1}
\end{equation}%
one obtains the leptonic observables, 
\begin{equation}
\begin{array}{l}
m_{e}=0.50\ \ \mathrm{MeV} \\ 
m_{\mu }=103.9\ \ \mathrm{MeV} \\ 
m_{\tau }=1.74\ \ \mathrm{GeV}%
\end{array}%
;\ \ 
\begin{array}{l}
\Delta m_{21}^{2}=8.8\ \ 10^{-5}\ \mathrm{eV}^{2} \\ 
\Delta m_{31}^{2}=3.1\ \ 10^{-3}\ \mathrm{eV}^{2} \\ 
m_{\upsilon 3}=0.047\ \mathrm{eV}%
\end{array}%
;\ \ 
\begin{array}{l}
\sin ^{2}(\theta _{12})=0.29 \\ 
\sin ^{2}(\theta _{23})=0.49\  \\ 
\sin ^{2}(\theta _{13})=0.016%
\end{array}
\label{lepexp1}
\end{equation}%
For the quarks, choosing for the $c_{i}^{Q}$, $c_{j}^{u}$, $c_{j}^{d}$ and $%
A^{u,d},B^{u,d}$ 
\begin{equation}
\begin{array}{lll}
c_{1}^{Q}=0.100 & c_{1}^{u}=0.365 & c_{1}^{d}=0.619 \\ 
c_{2}^{Q}=0.475 & c_{2}^{u}=0.655 & c_{2}^{d}=0.652 \\ 
c_{3}^{Q}=0.100 & c_{3}^{u}=0.448 & c_{3}^{d}=0.731%
\end{array}%
;\ \ \ 
\begin{array}{ll}
A^{u}=1.0072\ e^{0.0035i} & A^{d}=1.105\ e^{-0.0171i} \\ 
&  \\ 
B^{u}=1.007\ e^{0.0035i} & B^{d}=1.035\ e^{-0.0171i}%
\end{array}
\label{qua1}
\end{equation}%
yields the following $V_{CKM}$ and masses (at $M_{Z}$), 
\begin{equation}
|V_{CKM}|=\left[ 
\begin{array}{ccc}
0.9755 & 0.2201 & 0.0036 \\ 
0.2202 & 0.9747 & 0.0381 \\ 
0.0081 & 0.0374 & 0.9993%
\end{array}%
\right] \ ;\ \ 
\begin{array}{ll}
m_{u}=1.1\ \ \mathrm{MeV} & m_{d}=2.7\ \ \mathrm{MeV} \\ 
m_{c}=651\ \ \mathrm{MeV} & m_{s}=83.1\ \ \mathrm{MeV} \\ 
m_{t}=185\ \ \mathrm{GeV} & m_{b}=3.1\ \ \mathrm{GeV}%
\end{array}%
\   \label{vckm1}
\end{equation}

We have allowed for CP violation in the quark sector by choosing just two
phases for complex $A^{u,d},B^{u,d}$ permutation parameters. As a result we
obtain for the CP violation parameter $J_{CP}\equiv |\mathrm{Im}\
(V_{12}V_{23}V_{22}^{\ast }V_{13}^{\ast })|=2.8~10^{-5}$. The experimental
values for the angles of the unitarity triangle $\beta \equiv \arg
(-V_{21}V_{33}V_{23}^{\ast }V_{31}^{\ast })$ and $\gamma \equiv \arg
(-V_{11}V_{23}V_{13}^{\ast }V_{21}^{\ast })$, which at this moment measure $%
\sin (2\beta )_{\exp }=0.687\pm 0.032$ and $\gamma _{\exp }=\left(
63_{-12}^{+15}\right) ^{o}$, are near the values found here: $\sin (2\beta
)=0.76$ and $\gamma =73.5^{o}$.

(B) \textbf{USY}

In the case of the Universal Strength of Yukawa couplings hypothesis,
choosing for the leptons%
\begin{equation}
\begin{array}{lll}
c_{1}^{L}=0.2 & c_{1}^{l}=0.729 & c_{1}^{\upsilon }=1.387 \\ 
c_{2}^{L}=0.2 & c_{2}^{l}=0.637 & c_{2}^{\upsilon }=1.423 \\ 
c_{3}^{L}=0.2 & c_{3}^{l}=0.684 & c_{3}^{\upsilon }=1.423%
\end{array}%
\ ;\ \ \ k^{\upsilon }=\left[ 
\begin{array}{ccc}
1 & 1 & 1 \\ 
1 & -i & 1 \\ 
1 & 1 & i%
\end{array}%
\right] ;\ \ \ k^{l}=\left[ 
\begin{array}{ccc}
1 & 1 & 1 \\ 
1 & e^{0.012i} & 1 \\ 
1 & 1 & e^{0.563i}%
\end{array}%
\right]  \label{lep2}
\end{equation}%
yields the leptonic observables, 
\begin{equation}
\begin{array}{l}
m_{e}=0.48\ \ \mathrm{MeV} \\ 
m_{\mu }=104.6\ \ \mathrm{MeV} \\ 
m_{\tau }=1.75\ \ \mathrm{GeV}%
\end{array}%
;\ \ 
\begin{array}{l}
\Delta m_{21}^{2}=8.0\ \ 10^{-5}\ \mathrm{eV}^{2} \\ 
\Delta m_{31}^{2}=2.4\ \ 10^{-3}\ \mathrm{eV}^{2} \\ 
m_{\upsilon 3}=0.050\ \mathrm{eV}%
\end{array}%
;\ \ 
\begin{array}{l}
\sin ^{2}(\theta _{12})=0.25 \\ 
\sin ^{2}(\theta _{23})=0.39\  \\ 
\sin ^{2}(\theta _{13})=0.026%
\end{array}
\label{lepexp2}
\end{equation}%
Note the peculiar phases in the Yukawa structure for the neutrinos. Taking
for the quarks, 
\begin{equation}
\begin{array}{l}
\begin{array}{lll}
c_{1}^{Q}=0.30 & c_{1}^{u}=0.466 & c_{1}^{d}=0.631 \\ 
c_{2}^{Q}=0.18 & c_{2}^{u}=0.462 & c_{2}^{d}=0.631 \\ 
c_{3}^{Q}=0.30 & c_{3}^{u}=0.376 & c_{3}^{d}=0.637%
\end{array}%
;\ \ \ k^{u}=\left[ 
\begin{array}{ccc}
1 & 1 & e^{-0.015911i} \\ 
1 & 1 & e^{-0.015i} \\ 
e^{-0.01591i} & e^{-0.015i} & e^{-0.015i}%
\end{array}%
\right] \\ 
\\ 
k^{d}=diag(1,e^{-0.002i},e^{-0.045i})\cdot \left[ 
\begin{array}{ccc}
1 & 1 & e^{0.147i} \\ 
1 & 1 & e^{0.175i} \\ 
e^{0.147i} & e^{0.175i} & e^{0.175i}%
\end{array}%
\right]%
\end{array}
\label{qua2}
\end{equation}%
results in the following $V_{CKM}$ and masses (at $M_{Z}$), 
\begin{equation}
|V_{CKM}|=\left[ 
\begin{array}{ccc}
0.9754 & 0.2202 & 0.0032 \\ 
0.2201 & 0.9747 & 0.0381 \\ 
0.0082 & 0.0374 & 0.9993%
\end{array}%
\right] \ ;\ \ 
\begin{array}{ll}
m_{u}=1.3\ \ \mathrm{MeV} & m_{d}=2.8\ \ \mathrm{MeV} \\ 
m_{c}=654\ \ \mathrm{MeV} & m_{s}=93.1\ \ \mathrm{MeV} \\ 
m_{t}=186\ \ \mathrm{GeV} & m_{b}=3.0\ \ \mathrm{GeV}%
\end{array}%
\   \label{vckm2}
\end{equation}

In the USY case, the Yukawa couplings are already complex. For the quarks
and the phases given above, one finds $J_{CP}=2.6~10^{-5}$ together with $%
\sin (2\beta )=0.69$ and $\gamma =76.1^{o}$.

\section{Conclusions}

\label{conclu}

We have introduced, in a RS model with SM fields in the bulk and the Higgs
boson on the TeV-brane, two specific structures for the Yukawa couplings,
one based on a permutation symmetry and the other on the Universal Strength
of Yukawa couplings hypothesis. In both scenarios, all Yukawa couplings are
of the same order of magnitude. Thus, the main features of the fermion
hierarchies are explained through the RS geometrical mechanism, and not
because some Yukawa coupling is extremely small.

We have found that the RS model is particularly appropriate to incorporate
the two suggested Yukawa configurations. Indeed, the RS geometrical
mechanism of fermion locations along the extra dimension, combined with the
two Yukawa scenarios, reproduces all the present experimental data on
fermion masses and mixing angles. In particular, we find it remarkable that
in the USY case, only two complex phases of definite value $\pm \frac{\pi }{2%
}$ are sufficient to generate the known neutrino mass differences while at
same time permitting large leptonic mixing in agreement with experiment. We
also point out that for this RS model, in the USY case for the quarks, the
values for the CP violation parameters are much inproved, in contrast with
the SM-USY where CP violation is small.

\noindent

\textbf{\Large Acknowledgments}

\noindent This work was supported by the Funda\c{c}\~{a}o para a Ci\^{e}ncia
e Tecnologia (FCT) of the Portuguese Ministry of Science and Technology, and
the EU, through the projects POCTI/FNU/44409/2002 and PDCT/FP/63914/2005.
The author is grateful to D. Emmanuel-Costa for useful discussions and to
the CERN Physics Theory (PH) Department for the warm hospitality.


\begin{thebibliography}{99}
\bibitem{Kaluza} T.~Kaluza, Sitzungsber. Preuss. Akad. Wiss. Berlin (Math.
Phys.) \textbf{1921} (1921) 966; O.~Klein, Z. Phys. \textbf{37} (1926) 895
[Surveys High Energ. Phys. \textbf{5} (1986) 241].

\bibitem{String} For a review of recent developments, see K.~R.~Dienes,
Phys. Rept. \textbf{287} (1997) 447.

\bibitem{UED} T.~Appelquist, H.-C.~Cheng and B.~A.~Dobrescu, Phys. Rev. 
\textbf{D64} (2001) 035002.

\bibitem{Akama} K.~Akama, in \textit{Gauge Theory and Gravitation},
Proceedings of the International Symposium, Nara, Japan, 1982, ed. by
K.~Kikhawa, N.~Nakanishi and H.~Nariai (Springer-Verlag, 1983), 267,
hep-th/0001113\texttt{; }M.~Visser, Phys. Lett. \textbf{B159} (1985) 22;
E.~J.~Squires, Phys. Lett. \textbf{B167} (1985) 286; I.~Antoniadis, Phys.
Lett. \textbf{B246} (1990) 377.

\bibitem{Rubakov} V.~A.~Rubakov and M.~E.~Shaposhnikov, Phys. Lett. \textbf{%
B125} (1983) 136.

\bibitem{IntermA} I.~Antoniadis, C.~Mu\~{n}oz and M.~Quir\'{o}s, Nucl. Phys. 
\textbf{B397} (1993) 515; I.~Antoniadis, K.~Benakli and M.~Quir\'{o}s, Phys.
Lett. \textbf{B331} (1994) 313; I.~Antoniadis, S.~Dimopoulos, A.~Pomarol and
M.~Quir\'{o}s, Nucl. Phys. \textbf{B544} (1999) 503.

\bibitem{ADD} N.~Arkani-Hamed, S.~Dimopoulos and G.~Dvali, Phys. Lett. 
\textbf{B429} (1998) 263; I.~Antoniadis, N.~Arkani-Hamed, S.~Dimopoulos and
G.~Dvali, Phys. Lett. \textbf{B436} (1998) 257; N.~Arkani-Hamed,
S.~Dimopoulos and G.~Dvali, Phys. Rev. \textbf{D59} (1999) 086004.

\bibitem{Gog} M.~Gogberashvili, Int. J. Mod. Phys. \textbf{D11} (2002) 1635%
\texttt{.}

\bibitem{RS} L.~Randall and R.~Sundrum, Phys. Rev. Lett. \textbf{83} (1999)
3370.

\bibitem{SUSY} For a pedagogical text, see e.g. D.~Bailin and A.~Love, 
\textit{\textquotedblleft Supersymmetric gauge field theory and string
theory\textquotedblright }, Graduate student series in physics, Institute of
physics publishing, ed. by Douglas F.~Brewer.

\bibitem{UNI-RS} A.~Pomarol, Phys. Rev. Lett. \textbf{85} (2000) 4004;
L.~Randall and M.~D.~Schwartz, JHEP \textbf{0111} (2001) 003; L.~Randall and
M.~D.~Schwartz, Phys. Rev. Lett. \textbf{88} (2002) 081801; W.~D.~Goldberger
and I.~Z.~Rothstein, Phys. Rev. \textbf{D68} (2003) 125011; K.~Agashe,
A.~Delgado and R.~Sundrum, Annals Phys. \textbf{304} (2003) 145.

\bibitem{UNI} K.~R.~Dienes, E.~Dudas and T.~Gherghetta, Nucl. Phys. \textbf{%
B537} (1999) 47; Y.~Nomura, D.~Smith and N.~Weiner, Nucl. Phys. \textbf{B613}
(2001) 147.

\bibitem{KKrelic} E.~W.~Kolb and R.~Slansky, Phys. Lett. \textbf{B135}
(1984) 378; G.~Servant and T.~M.~P.~Tait, Nucl. Phys. \textbf{B650} (2003)
391, and references therein; H.-C.~Cheng, J.~L.~Feng and K.~T.~Matchev,
Phys. Rev. Lett. \textbf{89} (2002) 211301; D.~Hooper and G.~D.~Kribs, Phys.
Rev. \textbf{D67} (2003) 055003; K.~Agashe and G.~Servant, JCAP \textbf{0502}
(2005) 002 ; K.~Agashe and G.~Servant, Phys. Rev. Lett. \textbf{93} (2004)
231805.

\bibitem{AS} N.~Arkani-Hamed and M.~Schmaltz, Phys. Rev. \textbf{D61} (2000)
033005; M.~V.~Libanov and S.~V.~Troitsky, Nucl. Phys. \textbf{B599} (2001)
319; J.-M.~Fr\`{e}re, M.~V.~Libanov and S.~V.~Troitsky, Phys. Lett. \textbf{%
B512} (2001) 169; J.-M.~Fr\`{e}re, M.~V.~Libanov and S.~V.~Troitsky, JHEP 
\textbf{0111} (2001) 025; M.~V.~Libanov and E.~Ya.~Nougaev, JHEP \textbf{0204%
} (2002) 055; G.~Dvali and M.~Shifman, Phys. Lett. \textbf{B475} (2000) 295;
P.~Q.~Hung, Phys. Rev. \textbf{D67} (2003) 095011; D.~E.~Kaplan and
T.~M.~P.~Tait, JHEP \textbf{0006} (2000) 020; D.~E.~Kaplan and
T.~M.~P.~Tait, JHEP \textbf{0111} (2001) 051; M.~Kakizaki and M.~Yamaguchi,
Prog. Theor. Phys. \textbf{107} (2002) 433; Int. J. Mod. Phys. \textbf{A19}
(2004) 1715\texttt{;\ }C.~V.~Chang \textit{et al.}, Phys. Lett. \textbf{B558}
(2003) 92; S.~Nussinov and R.~Shrock, Phys. Lett. \textbf{B526} (2002) 137.

\bibitem{hep-ph/9912265} E.~A.~Mirabelli and M.~Schmaltz, Phys. Rev. \textbf{%
D61} (2000) 113011.

\bibitem{Bare} G.~Barenboim, G.~C.~Branco, A.~de Gouv\^{e}a and
M.~N.~Rebelo, Phys. Rev. \textbf{D64} (2001) 073005; G.~C.~Branco, A.~de Gouv%
\^{e}a and M.~N.~Rebelo, Phys. Lett. \textbf{B506} (2001) 115; P.~Q.~Hung
and M.~Seco, Nucl. Phys. \textbf{B653} (2003) 123.

\bibitem{Klap} H.~V.~Klapdor-Kleingrothaus and U.~Sarkar, Phys. Lett. 
\textbf{B541} (2002) 332; M.~Raidal and A.~Strumia, Phys. Lett. \textbf{B553}
(2003) 72; J.-M.~Fr\`{e}re, G.~Moreau and E.~Nezri, Phys. Rev. \textbf{D69}
(2004) 033003.

\bibitem{RSloc} T.~Gherghetta and A.~Pomarol, Nucl. Phys. \textbf{B586}
(2000) 141.

\bibitem{RSmany} D.~Dooling and K.~Kang, Phys. Lett. \textbf{B502} (2001)
189.

\bibitem{FN} C.~D.~Froggatt and H.~B.~Nielsen, Nucl. Phys. \textbf{B147}
(1979) 277.

\bibitem{others1} K.~R.~Dienes, E.~Dudas and T.~Gherghetta, Phys. Lett. 
\textbf{B436} (1998) 55; K.~R.~Dienes, E.~Dudas and T.~Gherghetta, Nucl.
Phys. \textbf{B537} (1999) 47; K.~Yoshioka, Mod. Phys. Lett. \textbf{A15}
(2000) 29; M.~Bando, T.~Kobayashi, T.~Noguchi and K.~Yoshioka, Phys. Rev. 
\textbf{D63} (2001) 113017; A.~Neronov, Phys. Rev. \textbf{D65} (2002)
044004; N.~Arkani-Hamed \textit{et al.}, Phys. Rev. \textbf{D61} (2000)
116003.

\bibitem{VolMesPow1} K.~R.~Dienes, E.~Dudas and T.~Gherghetta, Nucl. Phys. 
\textbf{B557} (1999) 25; N.~Arkani-Hamed \textit{et al.}, Phys. Rev. \textbf{%
D65} (2002) 024032; N.~Arkani-Hamed and S.~Dimopoulos, Phys. Rev. \textbf{D65%
} (2002) 052003.

\bibitem{GNeubertA} Y.~Grossman and M.~Neubert, Phys. Lett. \textbf{B474}
(2000) 361.

\bibitem{GNeubertB} T.~Appelquist \textit{et al.}, Phys. Rev. \textbf{D65}
(2002) 105019; T.~Gherghetta, Phys. Rev. Lett. \textbf{92} (2004) 161601.

\bibitem{mypaper} G.~Moreau, Eur. Phys. J. \textbf{C40} (2005) 539.

\bibitem{Eigencite1} S. Chang, C.S. Kim and M. Yamaguchi, Phys. Rev. \textbf{%
D73} (2006) 033002; K.R. Dienes and S. Hossenfelder, Phys. Rev. \textbf{D74}
(2006) 065013; J.D. Lykken, hep-ph/0607149; C. Biggio, F. Feruglio, I.
Masina and M. Perez-Victoria, Nucl. Phys. \textbf{B677 (}2004)\ 451.

\bibitem{HSHHLL} S.~J.~Huber and Q.~Shafi, Phys. Lett. \textbf{B544} (2002)
295; S.~J.~Huber and Q.~Shafi, Phys. Lett. \textbf{B583} (2004) 293.

\bibitem{HSpre} S.~J.~Huber and Q.~Shafi, Phys. Lett. \textbf{B512} (2001)
365.

\bibitem{Eigen} G. Moreau and J.I. Silva-Marcos, JHEP \textbf{0601} (2006)
048.

\bibitem{Eigen1} G. Moreau and J.I. Silva-Marcos, JHEP \textbf{0603} (2006)
090.

\bibitem{USY} G.C. Branco, J.I. Silva-Marcos and M.N. Rebelo, Phys. Lett. 
\textbf{B237} (1990) 446; Phys. Lett. \textbf{B428} (1998)136; Phys. Rev. 
\textbf{D62} (2000) 073004; Nucl. Phys. \textbf{B686} (2204) 188; G.C.
Branco and J.I. Silva-Marcos, Phys. Lett. \textbf{B359} (1995) 166; G.C.
Branco, D. Emmanuel-Costa, J.I. Silva-Marcos, Phys. Rev. \textbf{D56} (1997)
107; J.I. Silva-Marcos, Phys .Rev. \textbf{D59} (1999) 091301;

\bibitem{+-+A} I.~I.~Kogan, S.~Mouslopoulos, A.~Papazoglou, G.~G.~Ross and
J.~Santiago, Nucl. Phys. \textbf{B584} (2000) 313; S.~Mouslopoulos and
A.~Papazoglou, JHEP \textbf{0011} (2000) 018; I.~I.~Kogan, S.~Mouslopoulos,
A.~Papazoglou and G.~G.~Ross, Nucl. Phys. \textbf{B595} (2001) 225;
N.~Kaloper, Phys. Rev. \textbf{D60} (1999) 123506; I.~I.~Kogan,
S.~Mouslopoulos and A.~Papazoglou, Phys. Lett. \textbf{B501} (2001) 140;
I.~Oda, Phys. Lett. \textbf{B480} (2000) 305; I.~Oda, Phys. Lett. \textbf{%
B472} (2000) 59; H.~Hatanaka \textit{et al.}, Prog. Theor. Phys. \textbf{102}
(1999) 1213; C.~Csaki and Y.~Shirman, Phys. Rev. \textbf{D61} (2000)
024008;~N.~Arkani-Hamed, S.~Dimopoulos, G.~Dvali and N.~Kaloper, Phys. Rev.
Lett. \textbf{84} (2000) 586; R.~Gregory, V.~A.~Rubakov and
S.~M.~Sibiryakov, Phys. Rev. Lett. \textbf{84} (2000) 5928; T.~Li, Phys.
Lett. \textbf{B478} (2000) 307; L.~Randall and R.~Sundrum, Phys. Rev. Lett. 
\textbf{83} (1999) 4690; J.~Lykken and L.~Randall, JHEP \textbf{0006} (2000)
014; N.~Kaloper, Phys. Lett. \textbf{B474} (2000) 269; S.~Nam, JHEP \textbf{%
0003} (2000) 005; S.~Nam, JHEP \textbf{0004} (2000) 002.

\bibitem{HRizzo} H.~Davoudiasl, J.~L.~Hewett and T.~G.~Rizzo, Phys. Lett. 
\textbf{B473} (2000) 43; A.~Pomarol, Phys. Lett. \textbf{B486} (2000) 153.

\bibitem{hep-ph/9912498} S.~Chang \textit{et al.}, Phys. Rev. \textbf{D62}
(2000) 084025 ; B.~Bajc and G.~Gabadadze, Phys. Lett. \textbf{B474} (2000)
282.

\bibitem{Rebbi} R.~Jackiw and C.~Rebbi, Phys. Rev. \textbf{D13} (1976) 3398.

\bibitem{Tamvakis} A.~Kehagias and K.~Tamvakis, Phys. Lett. \textbf{B504}
(2001) 38.

\bibitem{GWise} W.~D.~Goldberger and M.~B.~Wise, Phys. Rev. Lett. \textbf{83}
(1999) 4922.

\bibitem{HSquark} S.~J.~Huber and Q.~Shafi, Phys. Lett. \textbf{B498} (2001)
256.

\bibitem{HSfv} S.~J.~Huber, Nucl. Phys. \textbf{B666} (2003) 269.

\bibitem{EWboundA} J.~L.~Hewett, F.~J.~Petriello and T.~G.~Rizzo, JHEP 
\textbf{0209} (2002) 030.

\bibitem{hep-ph/0204002} C.~S.~Kim, J.~D.~Kim and J.~Song, Phys. Rev. 
\textbf{D67} (2003) 015001.

\bibitem{EWboundB} S.~J.~Huber and Q.~Shafi, Phys. Rev. \textbf{D63} (2001)
045010.

\bibitem{EWboundC} S.~J.~Huber, C.-A.~Lee and Q.~Shafi, Phys. Lett. \textbf{%
B531} (2002) 112.

\bibitem{EWboundD} H.~Davoudiasl, J.~L.~Hewett and T.~G.~Rizzo, Phys. Rev. 
\textbf{D63} (2001) 075004.

\bibitem{EWboundE} G.~Burdman, Phys. Rev. \textbf{D66} (2002) 076003.

\bibitem{Eigencite2} E. De Pree, M. Sher, Phys. Rev. \textbf{D73} (2006)
095006; K. Agashe, A.E. Blechman and F. Petriello, Phys. Rev. \textbf{D74}
(2006) 053011; K. Agashe, G. Perez and A. Soni, hep-ph/0606293; S. Chang ,
C.S. Kim and J. Song, hep-ph/0607313; A. Djouadi, G. Moreau and F. Richard,
hep-ph/0610173.

\bibitem{FCbound} F.~del~Aguila and J.~Santiago, Phys. Lett. \textbf{B493}
(2000) 175.

\bibitem{hep-ph/0303232} P.~Huber \textit{et al.}, Nucl. Phys. \textbf{B665}
(2003) 487.

\bibitem{Valle} M.~Maltoni, T.~Schwetz, M.~A.~Tortola, J.~W.~F.~Valle, New
J. Phys. \textbf{6} (2004) 122, hep-ph/0405172.

\bibitem{PDG} Particle Data Group, W.-M. Yao \textit{et al.}, J. Phys. 
\textbf{G33} (2006) 1.

\bibitem{Fus} H.~Fusaoka and Y.~Koide, Phys. Rev. \textbf{D57} (1998) 3986.

\bibitem{Farzan} Y.~Farzan, O.~L.~G.~Peres and A.~Yu.~Smirnov, Nucl. Phys. 
\textbf{B612} (2001) 59.

\bibitem{Troitsk} V.~M.~Lobashev \textit{et al.}, Phys. Lett. \textbf{B460}
(1999) 227; Nucl. Phys. Proc. Suppl. \textbf{77} (1999) 327; \textbf{91}
(2001) 280; J.~Bonn \textit{et al.}, Nucl. Phys. Proc. Suppl. \textbf{91}
(2001) 273.
\end{thebibliography}
\end{document}